# Complexity of Manipulating Elections with Few Candidates

**Vincent Conitzer** and **Tuomas Sandholm**
{conitzer, sandholm}@cs.cmu.edu

**Abstract**

In multiagent settings where the agents have different preferences, preference aggregation is a central issue. Voting is a general method for preference aggregation, but seminal results have shown that all general voting protocols are manipulable. One could try to avoid manipulation by using voting protocols where determining a beneficial manipulation is hard. Especially among computational agents, it is reasonable to measure this hardness by computational complexity. Some earlier work has been done in this area, but it was assumed that the number of voters and candidates is unbounded. We derive hardness results for practical multiagent settings where the number of candidates is small but the number of voters can be large. We show that with complete information about the others' votes, individual manipulation is easy, and coalitional manipulation is easy with unweighted voters. However, constructive coalitional manipulation with weighted voters is intractable for all of the voting protocols under study, except for the nonrandomized Cup. Destructive manipulation tends to be easier. Randomizing over instantiations of the protocols (such as schedules of the Cup protocol) can be used to make manipulation hard. Finally, we show that under weak assumptions, if weighted coalitional manipulation with complete information about the others' votes is hard in some voting protocol, then individual and unweighted manipulation is hard when there is uncertainty about the others' votes.

## 1. Introduction

In multiagent settings, agents generally have different preferences, and it is of central importance to be able to aggregate these, i.e. to pick a socially desirable *candidate* from a set of candidates. Such candidates could be potential presidents, joint plans, allocations of goods or resources, etc. Voting is the most general preference aggregation scheme, and voting mechanisms have been used also for computational agents (e.g. (Ephrati & Rosenschein 1991; Ephrati & Rosenschein 1993)).

One key problem voting schemes are confronted with is that of *manipulation* by the voters. An agent is said to vote strategically when it does not rank the alternatives according to its true preferences, but rather so as to make the eventual outcome most favorable to itself. For example, if an agent prefers Nader to Gore to Bush, but knows that Nader has too few other supporters to win, while Gore and Bush are close to each other, the agent would be better off by declaring Gore as its top candidate. Manipulation is an undesirable phenomenon because social choice schemes are tailored to aggregate preferences in a socially desirable way, and if the agents reveal their preferences insincerely, a socially undesirable candidate may be chosen.

The issue of strategic voting has been studied extensively. A seminal negative result, the *Gibbard-Satterthwaite theorem*, shows that if there are three or more candidates, then in any nondictatorial voting scheme, there are preferences under which an agent is better off voting strategically (Gibbard 1973; Satterthwaite 1975). (A voting scheme is called dictatorial if one of the voters dictates the social choice no matter how the others vote).

When the voters are software agents, the algorithms they use to decide how to vote must be coded explicitly. Given that the voting algorithm needs to be designed only once (by an expert), and can be copied to large numbers of agents (even ones representing unsophisticated human voters), it is likely that rational strategic voting will increasingly become an issue, unmuddied by irrationality, emotions, etc.

Especially in the context of software agents, it is meaningful to ask how complex it is to compute a beneficial manipulation. It is conceivable that *complexity can be used as a desirable property* because if it is too complex to manipulate, the voters will not be able to do so. *Designing voting protocols where manipulation is complex promises to be an avenue for circumventing the fundamental impossibility results regarding the existence of non-manipulable voting protocols.*

The computational complexity of manipulation has already received some attention (Bartholdi, Tovey, & Trick 1989a; Bartholdi & Orlin 1991). However, the results to date that do show high complexity rely on both the number of candidates and the number of voters being unbounded. In most practical voting settings, the number of candidates is relatively small although the number of voters can be large. In this paper we show high complexity results for the practical setting where the number of candidates is finite (even just a small constant), and only the number of voters grows. Furthermore, we show low complexity in settings where both the number of candidates and voters are unbounded.

Restricting the number of candidates to a constant reduces the number of possible votes for a single voter to a constant. If the voters all have equal weight in the election, even the number of *de facto* possible combinations of votes that a coalition can submit is polynomial in the number of voters in the coalition (since the voters have equal weight, it does not matter which agent in the coalition submitted which vote; only the multiplicities of the votes from the coalition matter).

We get the following straightforward result.

**Proposition 1** *Let there be a finite number of candidates, and suppose that evaluating the result of a particular combination of votes by a coalition is in $\mathcal{P}$. If there is only one voter in the coalition, or if the voters are unweighted, the manipulation problem is in $\mathcal{P}$. (This holds for all the different variants of the manipulation problem, discussed later.)*

**Proof**: The manipulators (an individual agent or a coalition) can simply enumerate and evaluate all possibilities (there is a polynomial number of them). ■

In particular, in the complete-information manipulation problem in which the votes of the non-colluders are known, evaluating the result of a (coalitional) vote is as easy as determining the winner of an election, which must be in $\mathcal{P}$ for practical voting mechanisms.[1] This leaves open two avenues for deriving high complexity results with few candidates. First, we may investigate the complete-information coalitional manipulation problem when voters have *different weights*. While many human elections are unweighted, the introduction of weights generalizes the usability of voting schemes, and can be particularly important in multiagent systems settings with very heterogenous agents. We study this with deterministic voting protocols in Section 3, and in Section 4 we show that using randomization in the voting protocols can further increase manipulation complexity. Second, we may ask whether there are reasonable settings where *evaluating* a manipulation is $\mathcal{NP}$-hard. For instance, if we merely have probability distributions on the non-colluders' votes, how does the complexity of determining the probability that a given candidate wins change? We study this in Section 5, and show how to convert the results from Section 3 into stronger claims in this setting. In particular, we remove the assumptions of multiple manipulators and weighted votes.

## 2. Review of common voting protocols

In this section we define an election and the common voting protocols that we analyze.

**Definition 1** *An* election *consists of a set of $m$ candidates; a set of $n$ voters (possibly weighted), who are each to provide a total order on the candidates; and a function from the set of all possible combinations of votes to the set of candidates, which determines the winner.*

Different voting protocols are distinguished by their winner determination functions. We now review the most common protocols in use, all of which will be discussed in this paper. (We define them in the case of unweighted votes; the winner determination functions for weighted votes are defined by re-interpreting a vote of weight $k$ as $k$ identical unweighted votes[2]. Whenever points are defined, the candidate with the most points wins.)

---

[1] Theoretical voting mechanisms exist where determining the winner is $\mathcal{NP}$-hard (Bartholdi, Tovey, & Trick 1989b).

[2] We thus assume that weights are integers. The results also apply to all rational weights because they can be converted to integers by multiplying all the weights by all the weights' denominators.

- *Borda.* For each voter, a candidate receives $m-1$ points if it is the voter's top choice, $m-2$ if it is the second choice, ..., 0 if it is the last.

- *Copeland (aka. Tournament).* Simulate a pairwise election for each pair of candidates in turn (in a pairwise election, a candidate wins if it is preferred over the other candidate by more than half of the voters). A candidate gets 1 point if it defeats an opponent, 0 points if it draws, and -1 points if it loses.

- *Maximin.* A candidate's score in a pairwise election is the number of voters that prefer it over the opponent. A candidate's number of points is the lowest score it gets in any pairwise election.

- *Single Transferable Vote (STV).* The winner determination process proceeds in rounds. In each round, a candidate's score is the number of voters that rank it highest among the remaining candidates, and the candidate with the lowest score drops out. The last remaining candidate wins. (A vote '"transfers" from its top remaining candidate to the next highest remaining candidate when the former drops out.)

- *Cup (aka. Binary Protocol).* There is a balanced binary tree with one leaf per candidate. Then, each non-leaf node is assigned the candidate that is the winner of the pairwise election of the node's children. The candidate assigned to the root wins.

The Cup protocol requires a method for assigning (*scheduling*) candidates to leaf nodes. For instance, this assignment can be given *ex ante* (the "regular" Cup protocol), or we can randomize uniformly over the assignment after the votes have been submitted (the Randomized Cup protocol).

The winner determination function is not defined on all possible combinations of votes in these protocols since the tie-breaking methods are not specified. For simplicity, we assume tie-breaking mechanisms are adversarial to the manipulator(s), but this assumption is easy to relax without affecting the results of this paper.

## 3. Complexity of coalitional manipulation with weighted voters

In this section we discuss the complexity of constructive manipulation (causing a candidate to win) and destructive manipulation (causing a candidate to not win). We represent the order of candidates in a vote as follows: $(a_1, a_2, ..., a_m)$. If a vote's weight is not specified, it is 1. All our $\mathcal{NP}$-hardness reductions, directly or indirectly, will be from the PARTITION problem.

**Definition 2** PARTITION. *We are given a set of integers $\{k_i\}_{1 \leq i \leq t}$ (possibly with multiplicities) summing to $2K$, and are asked whether a subset of these integers sum to $K$.*

### 3.1. Constructive manipulation

**Definition 3** CONSTRUCTIVE-COALITIONAL-WEIGHTED-MANIPULATION (CCWM). *We are given*

*a set of weighted votes $S$, the weights for a set of votes $T$ which are still open, and a preferred candidate $p$. We are asked whether there is a way to cast the votes in $T$ so that $p$ wins the election.*[3]

**Theorem 1** *In the Borda protocol, CCWM is $\mathcal{N}P$-complete even with 3 candidates.*[4]

**Proof**: We reduce an arbitrary PARTITION instance to the following CCWM instance. There are 3 candidates, $a$, $b$ and $p$. In $S$ there are $6K-1$ voters voting $(a,b,p)$ and another $6K-1$ voting $(b,a,p)$. In $T$, for every $k_i$ there is a vote of weight $6k_i$. We show the instances are equivalent. Suppose there is a partition. Let the votes in $T$ corresponding to the $k_i$ in one half of the partition vote $(p,a,b)$, and let the other ones vote $(p,b,a)$. Then $a$ and $b$ each have $24K-3$ points, and $p$ has $24K$ points. So there is a manipulation. Conversely, suppose there is a manipulation. Then, since moving $p$ to the top of each vote in $T$ will never hurt $p$ in this protocol, there must exist a manipulation in which the only votes made in $T$ are $(p,a,b)$ and $(p,b,a)$. In this manipulation, since $p$ has $24K$ points in total and $a$ and $b$ each have $18K-3$ points from the votes in $S$, it follows that $a$ and $b$ each can gain at most $6K+2$ points from the votes in $T$. It follows that the voters voting $(p,a,b)$ can have combined total weight at most $6K+2$; hence the corresponding $k_i$ can sum to at most $K+\frac{1}{3}$, or (equivalently) to at most $K$ since the $k_i$ are all integers. The same holds for the $k_i$ corresponding to the $(p,b,a)$ votes. Hence, in both cases, they must sum to exactly $K$. But then, this is a partition. ∎

**Theorem 2** *In the Copeland protocol, CCWM is $\mathcal{N}P$-complete even with 4 candidates.*

**Proof**: We reduce an arbitrary PARTITION instance to the following CCWM instance. There are 4 candidates, $a$, $b$, $c$ and $p$. In $S$ there are $2K+2$ voters voting $(p,a,b,c)$, $2K+2$ voting $(c,p,b,a)$, $K+1$ voting $(a,b,c,p)$, and $K+1$ voting $(b,a,c,p)$. In $T$, for every $k_i$ there is a vote of weight $k_i$. We show the instances are equivalent. First, every pairwise election is already determined without $T$, except for the one between $a$ and $b$. $p$ defeats $a$ and $b$; $a$ and $b$ each defeat $c$; $c$ defeats $p$. If there is a winner in the pairwise election between $a$ and $b$, that winner will tie with $p$. So $p$ wins the Copeland election if and only if $a$ and $b$ tie in their pairwise election. But, after the votes in $S$ alone, $a$ and $b$ are tied. Thus, the votes in $T$ maintain this tie if and only if the combined weight of the votes in $T$ preferring $a$ to $b$ is the same as the combined weight of the votes in $T$ preferring $b$ to $a$. But this can happen if and only if there is a partition. ∎

**Theorem 3** *In the Maximin protocol, CCWM is $\mathcal{N}P$-complete even with 4 candidates.*

**Proof**: We reduce an arbitrary PARTITION instance to the following CCWM instance. There are 4 candidates, $a$, $b$, $c$ and $p$. In $S$ there are $7K-1$ voters voting $(a,b,c,p)$, $7K-1$ voting $(b,c,a,p)$, $4K-1$ voting $(c,a,b,p)$, and $5K$ voting $(p,c,a,b)$. In $T$, for every $k_i$ there is a vote of weight $2k_i$. We show the instances are equivalent. Suppose there is a partition. Then, let the votes in $T$ corresponding to the $k_i$ in one half of the partition vote $(p,a,b,c)$, and let the other ones vote $(p,b,c,a)$. Then, $p$ does equally well in each pairwise election: it always gets $9K$ pairwise points. $a$'s worst pairwise election is against $c$, getting $9K-1$. $b$'s worst is against $a$, getting $9K-1$. Finally $c$'s worst is against $b$, getting $9K-1$. Hence, $p$ wins the election. So there is a manipulation. Conversely, suppose there is a manipulation. Then, since moving $p$ to the top of each vote in $T$ will never hurt $p$ in this protocol, there must exist a manipulation in which all the votes in $T$ put $p$ at the top, and $p$ thus gets $9K$ as its worst pairwise score. Also, the votes in $T$ cannot change which each other candidate's worst pairwise election is: $a$'s worst is against $c$, $b$'s worst is against $a$, and $c$'s worst is against $b$. Since $c$ already has $9K-1$ points in its pairwise election against $b$, no vote in $T$ can put $c$ ahead of $b$. Additionally, if any vote in $T$ puts $a$ right above $c$, swapping their positions has no effect other than to decrease $a$'s final score, so we may also assume this does not occur. Similarly we can show it safe to also assume no vote in $T$ puts $b$ right above $a$. Combining all of this, we may assume that all the votes in $T$ vote either $(p,a,b,c)$ or $(p,b,c,a)$. Since $a$ already has $7K-1$ points in the pairwise election against $c$, the votes in $T$ of the first kind can have a total weight of at most $2K$; hence the corresponding $k_i$ can sum to at most $K$. The same holds for the $k_i$ corresponding to the second kind of vote on the basis of $b$'s score. Hence, in both cases, they must sum to exactly $K$. But then, this is a partition. ∎

**Theorem 4** *In the STV protocol, CCWM is $\mathcal{N}P$-complete even with 3 candidates.*

**Proof**: We reduce an arbitrary PARTITION instance to the following CCWM instance. There are 3 candidates, $a$, $b$ and $p$. In $S$ there are $6K-1$ voters voting $(b,p,a)$, $4K$ voting $(a,b,p)$, and $4K$ voting $(p,a,b)$. In $T$, for every $k_i$ there is a vote of weight $2k_i$. We show the instances are equivalent. Suppose there is a partition. Then, let the votes in $T$ corresponding to the $k_i$ in one half of the partition vote $(a,p,b)$, and let the other ones vote $(p,a,b)$. Then in the first round, $b$ has $6K-1$ points, $a$ has $6K$, and $p$ has $6K$. So $b$ drops out; all its votes transfer to $p$, so that $p$ wins the final round. So there is a manipulation. Conversely, suppose there is a manipulation. Clearly, $p$ cannot drop out in the first round; but also, $a$ cannot drop out in the first round, since all its votes in $S$ would transfer to $b$, and $b$ would have at least $10K-1$ points in the final round, enough to guarantee it victory. So, $b$ must drop out in the first round. Hence, from the votes in $T$, both $a$ and $c$ must get at least $2K$ weight that puts them in the top spot. The corresponding $k_i$ in either case must thus

---

[3]To an economist, it would be more natural to define a successful manipulation as one that increases the voter's (expected) utility. It is easy to see that our definitions are special cases of this utility-based definition, so our hardness results apply to that as well.

[4]In all $\mathcal{N}P$-completeness proofs, we only prove $\mathcal{N}P$-hardness because proving that the problem is in $\mathcal{N}P$ is trivial.

sum to at least $K$. Hence, in both cases, they must sum to exactly $K$. But then, this is a partition. ∎

### 3.2. Destructive manipulation

In the *destructive* version of the CCWM problem (which we call DCWM), instead of being asked whether there is a coalitional vote that makes $p$ win, we are asked whether there is a coalitional vote that makes $h$ not win. It is easy to see that DCWM can never be harder than CCWM (except by a factor $m$) because in order to solve DCWM we may simply solve CCWM once for each candidate besides $h$.

Interestingly, in these protocols (except STV), destructive manipulation turns out to be drastically easier than constructive manipulation!

**Theorem 5** *Consider any voting protocol where each candidate receives a numerical score based on the votes, and the candidate with the highest score wins. Suppose that the score function is monotone, that is, if voter $i$ changes its vote so that $\{b : a >_i^{old} b\} \subseteq \{b : a >_i^{new} b\}$, $a$'s score will not decrease. Finally, assume that the winner can be determined in polynomial time. Then for this protocol, DCWM is in $\mathcal{P}$.*

**Proof**: Consider the following algorithm: for each candidate $a$ besides $h$, we determine what the outcome of the election would be for the following coalitional vote. All the colluders place $a$ at the top of their votes, $h$ at the bottom, and order the other candidates in whichever way. We claim there is a vote for the colluders with which $h$ does not win if and only if $h$ does not win in one of these $m - 1$ elections. The *if* part is trivial. For the *only if* part, suppose there is a coalitional vote that makes $a \neq h$ win the election. Then, in the coalitional vote we examine where $a$ is always placed on top and $h$ always at the bottom, by monotonicity, $a$'s score cannot be lower and $h$'s cannot be higher than in the successful coalitional vote. It follows that here, too, $a$'s score is higher than $h$'s, and hence $h$ does not win the election. The algorithm is clearly in $\mathcal{P}$ since we do $m - 1$ winner determinations, and winner determination is in $\mathcal{P}$. ∎

**Corollary 1** *DCWM is in $\mathcal{P}$ for the Borda, Copeland, and Maximin protocols.*

The theorem does not apply to STV, however. We show that in fact, in STV, DCWM is $\mathcal{NP}$-complete.

**Theorem 6** *In the STV protocol, DCWM is $\mathcal{NP}$-complete even with 4 candidates.*

**Proof**: We reduce an arbitrary instance of CCWM for STV with 3 candidates to the following DCWM instance. Let the candidates in the original instance be $a$, $b$, and $p$; and let the voters (including the colluders) here have a combined weight of $W$. In the DCWM instance we construct, we have the same candidates plus an additional candidate $h$. $T$ (the number of colluders and their weights) remains exactly the same. $S$ includes all the votes (with weights) from the CCWM's instance's $S$ ($h$ is added to the bottom of these votes); additionally, there are the following 6 votes: $(a, b, p, h)$, $(a, p, b, h)$, $(b, a, p, h)$, $(b, p, a, h)$, and two times $(p, h, a, b)$. Finally, there are $W + 5$ votes voting $(h, a, b, p)$. We observe the following facts. First, $h$ will not be eliminated before the last round as it has close to half the vote weight at the start. Second, $h$ will lose this last round if and only if it faces $p$ in it, and all colluders have ranked $p$ above $h$. (If even one more vote transfers to $h$, it is certain to win the election as it has more than half the vote weight; and if $p$ drops out, the $(p, h, a, b)$ votes will transfer to $h$. On the other hand, $p$ is ranked above $h$ in all votes in $S$ that do not have $h$ at the top, so while $p$ remains none of these transfer to $h$.) Third, since the transfer of any additional vote to $h$ leads to its victory, all colluders may as well place $h$ at the bottom of their votes. Fourth, as long as $p$ has not dropped out, the relative scores of (the remaining candidates among) $a$, $b$, and $p$ in each round before the last will be exactly the same as in the CCWM instance if the coalition votes the same (disregarding $h$) in both instances. (The 6 additional votes in $S$ are carefully chosen to always be distributed equally among them while $p$ remains.) Thus, there is a coalitional vote that leads $p$ to the last round if and only if the CCWM instance has a constructive manipulation. Hence, by our second observation, the instances are equivalent. ∎

## 4. Increasing complexity via randomization

In this section, we investigate the effect of randomizing over different instantiations of a protocol on manipulation complexity. While most protocols only have one instantiation, the Cup protocol requires a schedule to be instantiated. We show that randomization over these schedules (after the votes have been cast) is sufficient to make manipulation $\mathcal{N}P$-complete. We first show that the Cup protocol is easy to manipulate if the schedule is known in advance.

**Theorem 7** *In the Cup protocol (given the assignment of candidates to leaves), CCWM is in $\mathcal{P}$.*

**Proof**: We demonstrate a method for finding all the potential winners of the election. In the binary tree representing the schedule, we can consider each node to be a subelection, and compute the set of potential winners for each subelection. (In such a subelection, we may say that the voters only order the candidates in that subelection since the place of the other candidates in the order is irrelevant.) Say a candidate *can* obtain a particular result in the election if it does so for some coalitional vote. The key claim to the proof, then, is the following: a candidate can win a subelection if and only if it can win one of its children, *and* it can defeat one of the potential winners of the sibling child in a pairwise election. It is easy to see that the condition is necessary. To show that it is sufficient, let $p$ be a candidate satisfying the condition by being able to defeat $h$, a potential winner of the other child (or *half*). Consider a coalitional vote that makes $p$ win its half, and another one that makes $h$ win its half. We now let each coalitional voter vote as follows: it ranks all the candidates in $p$'s half above all those in $h$'s half; the rest of the order is the same as in the votes that make $p$ and $h$ win their halves. Clearly, this will make $p$ and $h$ the finalists. Also, $p$ will win the pairwise election against $h$ since it is always ranked above $h$ with the colluders; and as we know that there is some coalitional vote that makes $p$ defeat $h$ pairwise, this one must have the same result. The obvious recursive algorithm has running time $O(m^3n)$ according to the Master Theorem (Cormen, Leiserson, & Rivest 1990). ∎

It turns out that randomizing over cup schedules makes manipulation hard even with few candidates, as the following definition and theorem show.

**Definition 4** UNCERTAIN-INSTANTIATION-CONSTRUCTIVE-COALITIONAL-WEIGHTED-MANIPULATION (UICCWM). *We are given a set of weighted votes $S$, the weights for a set of votes $T$ which are still open, a preferred candidate $p$, a distribution over instantiations of the voting protocol, and a number $r$, where $0 \leq r \leq 1$. We are asked whether there is a way to cast the votes in $T$ so that $p$ wins with probability greater than $r$.*

**Theorem 8** *In the Randomized Cup protocol, UICCWM is $\mathcal{NP}$-complete with 7 candidates.*

Because of limited space, we only sketch the proof.

**Sketch of Proof**: Given weights for the colluders (corresponding to the $k_i$ of a PARTITION instance), it is possible to define votes in $S$ over the 7 candidates ($a$ through $f$, and $p$) with the following properties. First, the colluders can only affect the outcomes of the pairwise elections between $a$, $b$, and $c$. Second, they can achieve the result that $a$ defeats $b$ (in their pairwise election), $b$ defeats $c$, and $c$ defeats $a$ if and only if their weights can be partitioned. Third, $d$ defeats $e$, $e$ defeats $f$, and $f$ defeats $d$. Fourth, $a$ defeats $d$, $b$ defeats $e$, and $c$ defeats $f$; otherwise, all of $d$, $e$, and $f$ defeat all of $a$, $b$, and $c$. Fifth, $p$ defeats all of $a$, $b$, and $c$, but loses to all of $d$, $e$, and $f$. Then it can be shown that if the colluders could decide each of the pairwise elections between $a$, $b$, and $c$ independently, letting $a$ defeat $b$, $b$ defeat $c$, and $c$ defeat $a$ strictly maximizes the probability that $p$ wins. (This is done by drawing the Cup tree (which has one bye round) and analyzing cases.) It follows that there exists a number $r$ ($0 \leq r \leq 1$) such that the colluders can make $p$ win with probability greater than $r$ if and only if there is a partition. ∎

## 5. Effect of uncertainty about others' votes

So far we discussed the complexity of coalitional manipulation when the others' votes are known. We now show how those results can be related to the complexity of manipulation by an individual voter when only a *distribution* over the others' votes is known. If we allow for arbitrary distributions, we need to specify a probability for each combination of votes by the others, that is, exponentially many probabilities (even with just two candidates). It is impractical to specify so many probabilities.[5] Therefore, it is reasonable to presume that the language used for specifying these probabilities would not be fully expressive. We derive the complexity results of this section for extremely restricted probability distributions, which any reasonable language should allow for. Thus our results apply to any reasonable language. Due to limited space, we only present results on constructive manipulations, but all results apply to the destructive cases as well and the proofs are analogous.

### 5.1. Weighted voters

First we show that with weighted voters, in protocols where coalitional manipulation is hard in the complete-information case, even evaluating a candidate's winning probability is hard when there is uncertainty about the votes (even when there is no manipulator).

**Definition 5** UNCERTAIN-VOTES-CONSTRUCTIVE-WEIGHTED-EVALUATION (UVCWE). *We are given a weight for each voter, a distribution over all the votes, a candidate $p$, and a number $r$, where $0 \leq r \leq 1$. We are asked whether the probability of $p$ winning is greater than $r$.*

**Theorem 9** *If CCWM is $\mathcal{NP}$-hard for a protocol (even with $k$ candidates), then UVCWE is also $\mathcal{NP}$-hard for it (even with $k$ candidates), even if the votes are drawn independently and only the following types of (marginal) distributions are allowed: 1) the vote's distribution is uniform over all possible votes, and 2) the vote's distribution puts all of the probability mass on a single vote.*

**Proof**: For the reduction from CCWM to UVCWE, we use exactly the same voters, and $p$ remains the same as well. If a voter was not a colluder in the CCWM instance and we were thus given its vote, in the UVCWE instance its distribution is degenerate at that vote. If the voter was in the collusion, its distribution is now uniform. We set $r = 0$. Now, clearly, in the PCWE instance there is a chance of $p$ winning if and only if there exists some way for the latter votes to be cast so as to make $p$ win - that is, if and only if there is an effective collusion in the CCWM problem. ∎

Next we show that if evaluating the winning probability is hard, individual manipulation is also hard.

**Definition 6** UNCERTAIN-VOTES-CONSTRUCTIVE-INDIVIDUAL-WEIGHTED-MANIPULATION (UVCIWM). *We are given a single manipulative voter with a weight, weights for all the other voters, a distribution over all the others' votes, a candidate $p$, and a number $r$, where $0 \leq r \leq 1$. We are asked whether the manipulator can cast its vote so that $p$ wins with probability greater than $r$.*

**Theorem 10** *If UVCWE is $\mathcal{NP}$-hard for a protocol (even with $k$ candidates and restrictions on the distribution), then UVCIWM is also $\mathcal{NP}$-hard for it (even with $k$ candidates and the same restrictions).*

**Proof**: For the reduction from UVCWE to UVCIWM, simply add a manipulator with weight 0. ∎

---

[5]Furthermore, if the input is exponential in the number of voters, an algorithm that is exponential in the number of voters is not necessarily complex in the usual sense of input complexity.

Combining Theorems 9 and 10, we find that with weighted voters, if in some protocol coalitional manipulation is hard in the complete-information setting, then even individual manipulation is hard if others' votes are uncertain. Applying this to the hardness results from Section 3, this means that all of the protocols of this paper other than Cup are hard to manipulate by individuals in the weighted case when the manipulator is uncertain about the others' votes.

## 5.2. Unweighted voters

Finally, we show that in protocols where coalitional manipulation is hard in the weighted complete-information case, evaluating a candidate's winning probability is hard even in the unweighted case when there is uncertainty about the votes (even when there is no manipulator). This assumes that the language for specifying the probability distribution is rich enough to allow for perfect correlations between votes (that is, some votes are identical with probability one[6]).

**Theorem 11** *If UVCWE is $\mathcal{N}P$-hard for a protocol (even with $k$ candidates and restrictions on the distribution), then the unweighted version of UVCWE is also $\mathcal{N}P$-hard for it if we allow for perfect correlations (even with $k$ candidates and the same restrictions—except those conflicting with perfect correlations).*

**Proof**: For the reduction from UVCWE to its unweighted version, we replace each vote of weight $k$ with $k$ unweighted votes; we then make these $k$ votes perfectly correlated. Subsequently we pick a representative vote from each perfectly correlated group, and we impose a joint distribution on these votes identical to the one on the corresponding votes in the UVCWE problem. This determines a joint distribution over all votes. It is easy to see that the distribution over outcomes is the same as in the instance we reduced from; hence, the decision questions are equivalent. ∎

## 6. Conclusions and future research

In multiagent settings where the agents have different preferences, preference aggregation is a central issue. Voting is a general method for preference aggregation, but seminal results have shown that all general voting protocols are manipulable. One could try to avoid manipulation by using voting protocols where determining a beneficial manipulation is hard. Especially among computational agents, it is reasonable to measure this hardness by computational complexity. Some earlier work has been done in this area, but it was assumed that the number of voters and candidates is unbounded.

In this paper we derived hardness results even in the practical case where the number of candidates is finite (a small constant). We showed that with complete information about the others' votes, individual manipulation is easy, and coalitional manipulation is easy with unweighted voters. However, constructive coalitional manipulation with weighted voters turned out to be intractable for all of the voting protocols under study, except for the nonrandomized Cup protocol where it is easy. Destructive manipulation tends to be easier. Randomizing over instantiations of the protocols (such as schedules of the Cup protocol) can be used to make manipulation hard. Finally, we showed that under weak assumptions, if weighted coalitional manipulation with complete information about the others' votes is hard in some voting protocol, then individual and unweighted manipulation is hard when there is uncertainty about the others' votes.

In summary, our results suggest preferring STV over other protocols on the basis of the difficulty of destructive manipulation; randomizing over various instantiations of a protocol; or, they may lead one to be generally less concerned about the possibility of manipulation as it appears difficult in the common case where not too much is known about how others will vote.

All of the results on complexity of manipulation to date (including ours) use $\mathcal{N}P$-hardness as the complexity measure. This is only a weak guarantee of hardness of manipulation. It means that there are infinitely many hard instances, but many (or even most) instances may be easy to manipulate. Future work includes studying other notions of hardness in the manipulation context, such as average case completeness (Gurevich 1991) and instance complexity (Orponen *et al.* 1994). Future work also includes proving hardness of manipulation in more restricted protocols such as auctions, and with more restricted preferences.

---

[6]Representation of such distributions can still be concise.